\documentclass[aps,prc,fleqn,floatfix,twocolumn,twoside]{revtex4}

\usepackage[dvipdfm]{graphicx}

\usepackage{amsmath,amssymb,amsfonts}

\usepackage{color}

\begin{document}

\title{Translation of collision geometry fluctuations into momentum anisotropies
in relativistic heavy-ion collisions}

\author{Guang-You Qin, Hannah Petersen, Steffen A. Bass and Berndt M\"uller}
\affiliation{Department of Physics, Duke University, Durham, NC 27708, USA}$$$$

\date{\today}
\begin{abstract}

We develop a systematic framework for the study of the initial collision
geometry fluctuations in relativistic heavy-ion collisions and investigate how
they evolve through different
stages of the fireball history and translate into final particle momentum
anisotropies.
We find in our event-by-event analysis that only the few lowest momentum
anisotropy parameters survive after the
hydrodynamical evolution of the system.
The geometry of the produced medium is found to be affected by 
the pre-equilibrium
evolution of the medium and the thermal smearing of the discretized
event-by-event initial conditions, both of which tend to smear out the spatial
anisotropies. We find such effects to be more prominent for higher moments than 
for lower moments.
The correlations between odd and even spatial anisotropy parameters during the
pre-equilibrium expansion are quantitatively studied and found to be
small.
Our study provides a theoretical foundation for the understanding of initial
state fluctuations and the collective
expansion dynamics in relativistic heavy-ion collisions.

\end{abstract}
\maketitle

\section{Introduction}

Experiments at the Relativistic Heavy Ion Collider (RHIC) have discovered that
the strongly interacting matter
produced in these highly energetic collisions exhibits strong collective flow,
which can be well described by relativistic
hydrodynamics \cite{Kolb:2000sd, Teaney:2000cw, Huovinen:2001cy, Hirano:2002ds, 
Huovinen:2005gy, Nonaka:2006yn, Schenke:2010nt, Niemi:2008ta}. In noncentral
collisions, the collective flow is
azimuthally asymmetric in the plane transverse to the beam axis. This has been
understood as the consequence of the
initial spatial asymmetry of the medium produced by the two colliding nuclei
which translates into a momentum
anisotropy of the emitted particles due to the hydrodynamic expansion of the
matter. The magnitude of this flow
anisotropy is quantified by the Fourier expansion coefficients $v_n$ of the
azimuthal angular distribution of the emitted
particles in the transverse plane \cite{Ollitrault:1992bk}.

The elliptic flow $v_2$ signal has been extensively studied in Au+Au collisions
at RHIC as a function of various
quantities \cite{Adams:2003zg}. Hydrodynamic simulations have shown that
elliptic flow $v_2$ is sensitive to various
transport properties of the expanding hot medium, especially the specific shear
viscosity $\eta$, the presence of
which tends to reduce the amount of the elliptic flow that can be built up in an
ideal hydrodynamical fluid
\cite{Teaney:2003kp, Romatschke:2007mq, Song:2007fn, Dusling:2007gi,Luzum:2008cw}.
Considerable effort has been devoted to
the quantitative extraction of the shear viscosity by comparing the
measured elliptic flow $v_2$ with viscous relativistic hydrodynamic simulation
of the fireball evolution
and other Boltzmann transport models that involve the violation of ideal
hydrodynamic behavior
\cite{Xu:2007jv, Drescher:2007cd}. These comparisons have yielded an upper limit
for the shear viscosity to entropy
density $s$ ratio: $\eta/s<0.5$ \cite{Song:2008hj, Nagle:2009ip}, the same order
of magnitude as the conjectured KSS
bound $\eta/s = 1/(4\pi)$ \cite {Kovtun:2004de}, which were obtained using
anti-de-Sitter/conformal field theory (AdS/CFT) correspondence for certain
quantum field
theories similar to QCD.

Current efforts in the extraction of the shear viscosity from precise $v_2$
measurements are subjected to various
uncertainties in the hydrodynamic simulations, i.e., equations of state, large
shear viscosity in late hadronic stage
\cite{Demir:2008tr}, bulk viscosity \cite{Song:2009rh}, and the treatment of the
freeze-out conditions.
Among the largest uncertainties is the initial geometry
employed in the hydrodynamical simulations, i.e., the initial fireball
eccentricity $\epsilon_2=\langle y^2-x^2\rangle
/ \langle y^2+x^2 \rangle$ \cite{Hirano:2005xf, Luzum:2008cw}. In ideal
hydrodynamics, the elliptic flow is built up
from pressure gradients and thus directly proportional to the initial fireball
eccentricity. Unfortunately, there has
been no direct experimental measurements of this quantity due to the difficulty
of isolating the initial state
contribution from the later stages of the fireball evolution.
Model estimates of the overlap geometry of two nuclei differ up to $20-30\%$ in
eccentricity, which turns out to
introduce more than a factor of two uncertainty in the extracted values for
$\eta/s$
\cite{Romatschke:2007mq}.
Therefore, the precise determination of $\eta/s$ requires a more precise
knowledge of the initial geometry for the produced fireball in the collisions.

Recently, significant attention has been paid to initial geometry fluctuations
\cite{Miller:2003kd, Broniowski:2007ft, Alver:2008zza, Hirano:2009bd,
Staig:2010pn, Mocsy:2010um} which have been used to explain
the underestimation of elliptic flow calculated in various ideal and viscous
hydrodynamic simulations for the most central
collisions. For example, the geometry fluctuations of the positions of nucleons
in the Monte Carlo Glauber (MCG) model
\cite{Glauber:1970jm, Miller:2007ri} lead to fluctuations of the participant
plane from one event to another,
rendering larger eccentricities which
translates into larger elliptic flow for the final state particles. To pursue
such studies, one needs to run hydrodynamical
evolution on an event-by-event basis utilizing fluctuating initial conditions
\cite{Petersen:2010md, Holopainen:2010gz,
Petersen:2010cw}.

As is known, higher order moments are also present in fluctuating initial
collision geometry when one performs a harmonic/multipole analysis.
Triangular geometry and flow have recently been proposed to explain features in
the data such as the ridge and broad away-side correlations observed in
two-particle
correlation data, in the context of  hydrodynamics and transport models
\cite{Alver:2010gr, Petersen:2010cw, Alver:2010dn}.
Higher-order flow coefficients have been measured \cite{Abelev:2007qg,
Adare:2010ux} and recent studies show that
the initial state density fluctuations may play an important role in
understanding the centrality dependence of the ratio $v_4/v_2^2$
\cite{Gombeaud:2009ye, Luzum:2010ae}. To achieve a full understanding of the
expansion dynamics of the produced
fireball therefore requires a systematic study of initial geometry fluctuations.
The main purpose of our paper is to investigate how
harmonic moments of different order propagate through the different stages of
the fireball history and how they translate themselves into the
momentum anisotropies of the final produced particles. 

In Section II, we construct the full phase space distribution of the initial
conditions (position and momentum space) obtained from a Monte Carlo Glauber
model with the inclusion of the nucleon position fluctuations as well as
fluctuations from individual nucleon-nucleon collisions. The geometry of such
initial conditions is analyzed in Section III.
We study the pre-equilibrium evolution of the system and its effect on the
spatial geometry in Section IV, where
a detailed analysis of the correlations between odd and even moments during this
period is also presented.
In Section V, the discretized initial conditions is smeared with a Gaussian
distribution and, assuming sudden thermalization, the subsequent evolution of
the system is modeled utilizing a three-dimensional relativistic ideal
hydrodynamics\cite{Rischke:1995ir, Rischke:1995mt, Petersen:2010cw}.
Numerical results of final state momentum anisotropies after the  hydrodynamical
evolution are presented in Section VI, followed by our summary in the last
section.

\section{Initial Conditions}

Our initial conditions are based on the  Monte Carlo Glauber model, but differ
from other implementations of that model as we include the
fluctuations of nucleon positions as well as the fluctuations originating from
individual nucleon-nucleon collisions. In addition we account for the full
phase-space
by constructing the particle momentum distributions as well.
We determine the spatial distribution using the well-established two-component
(binary collision and participant)
scaling and the momentum distribution is obtained by fitting to data on final
particle momentum spectra. We also treat the early pre-equilibrium
expansion of the system using the free-streaming approximation prior to the
hydrodynamical evolution.

We start with the nuclear distribution function inside a nucleus taken as the
Woods-Saxon form
\begin{align}
\rho_A(r) = \frac{\rho_0}{1 + \exp[(r-R)/d]}
\end{align}
where the radius $R$ and the diffuse constant $d$ are taken as $R=6.38~{\rm
fm}$, $d=0.535~{\rm fm}$ for a Au nucleus.
The above distribution is normalized to the atom number $\int d^3r \rho(r) = A$
with $\rho_0=0.163/{\rm fm}^3$.
It is convenient to normalize the above distribution function to unity and
define the
single nucleon distribution $\hat{\rho}_A(r)$ inside a nucleus, $\int d^3r
\hat{\rho}_A(r) = 1$. The normalized
thickness function is defined as
\begin{align}
\hat{T}_A({\bf s}) = \int dz \hat{\rho}_A({\bf s}, z)
\end{align}
with the normalization $ \int d^2{\bf s}\hat{T}_A({\bf s}) = 1$.

To study the collision between two incoming nuclei at a given impact parameter
${\bf b}$, one may define the probability for
a given nucleon $i$ from  nucleus $A$ and a given nucleon $j$ from  nucleus $B$
to collide to be $P({\bf s}_i, {\bf s}_j,
{\bf b}) = \hat{\sigma}({\bf s}_i-{\bf s}_j-{\bf b})$, which is normalized to
the nucleon-nucleon inelastic cross section
$\sigma_{NN}$,
\begin{align}
\int  d^2{\bf s} \hat{\sigma}({\bf s}) = \sigma_{NN}
\end{align}
where $\sigma_{NN}=42~{\rm mb}$ is taken for nucleon-nucleon collisions at
$\sqrt{s_{NN}}=200~{\rm GeV}$. From such a
probability distribution, one may compute the numbers of binary nucleon-nucleon
collisions and participating nucleons,
\begin{eqnarray}
N_{\rm coll} \!\!&&\!\!= \sum_{i=1}^A \sum_{j=1}^B \int d^2{\bf s}_i
\hat{T}_A({\bf s}_i) \int d^2{\bf s}_j
\hat{T}_B({\bf s}_j) \hat{\sigma}({\bf s}) \ \ \ \ \ \ \ \ \ \ \ \
\nonumber \\
 N_{\rm part} \!\!&&\!\!=  \sum_{i=1}^A  \int d^2{\bf s}_i \hat{T}_A({\bf s}_i)
\nonumber\\ &&   \left\{1 - \prod_{j=1}^B \int d^2{\bf
s}_j \hat{T}_B({\bf s}_j) [1 - \hat{\sigma}({\bf s})]\right\} + (A \leftrightarrow B)
\end{eqnarray}
where  ${\bf s} = {\bf s}_i-{\bf s}_j-{\bf b}$.

To simulate a collision of two nuclei using the Monte Carlo approach, one first
samples the positions of all nucleons in
the nucleus according to a Woods-Saxon distribution, and obtains discrete
nucleon distributions with each single nucleon
corresponding to a $\delta$ function.
The probability function $\hat{\sigma}({\bf s}_i - {\bf s}_j - {\bf b})$ for two
nucleons to collide is taken to be
geometrical in form
\begin{eqnarray}
&\!\!\hat{\sigma}({\bf s}_i-{\bf s}_j-{\bf b}) = 1 \,,&\,\,\, |{\bf s}_i - {\bf s}_j -
{\bf b}| \le \sqrt{\sigma_{NN}/\pi} \nonumber\\
&\!\!\hat{\sigma}({\bf s}_i-{\bf s}_j-{\bf b}) = 0 \,,&\,\,\, |{\bf s}_i - {\bf s}_j -
{\bf b}| > \sqrt{\sigma_{NN}/\pi} \ \
\
\end{eqnarray}
With this, one returns to the classical picture of collisions: two nucleons with
transverse distance $d_\perp = |{\bf s}_i -
{\bf s}_j - {\bf b}| \le \sqrt{\sigma_{NN}/\pi}$ will collide with each other.
Note that the assumption of linear trajectories of participant nucleons is still
maintained after they collide with each other.
With the above probability distribution, one reduces the calculation of $N_{\rm
coll}$ and $N_{\rm part}$ to counting the pairs of binary collisions and the
number of participating nucleons.

\begin{figure}[htb]
\includegraphics[width=8cm]{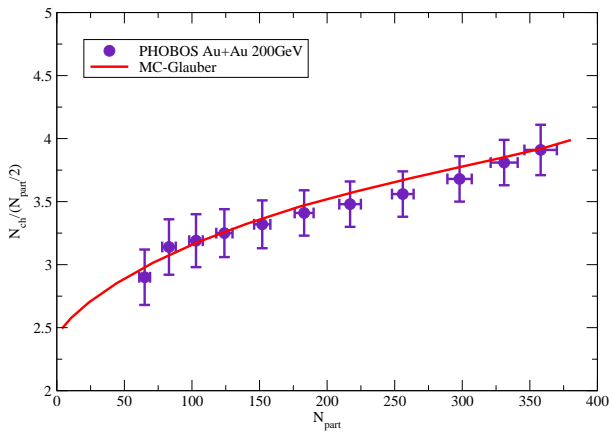}
 \caption{(Color online) Charged particle multiplicity at midrapidity
$|\eta|<0.5$ as a function of centrality in Au+Au collisions
 at $\sqrt{s_{NN}}=200$~GeV.
} \label{ncharge_npart}
\end{figure}

After determining the profiles of two colliding nuclei, the produced particle
multiplicity in a collisions for a given
centrality class (or impact parameter ${\rm b}$) and rapidity range $\Delta
\eta$ can be obtained from the following
phenomenological two-component formula \cite{Kharzeev:2000ph},
\begin{eqnarray}\label{2comp_nptc}
N_{AA}({\bf b}, \Delta \eta)  =  \left[\alpha  N_{\rm coll}({\bf b})  +
\frac{1\!-\!\alpha}{2} {N_{\rm part}({\bf b}}) \right]
N_{NN}(\Delta \eta)  \nonumber\\
\end{eqnarray}
where $N_{NN}(\Delta \eta)$ is the particle multiplicity in a nucleon-nucleon
collision at the same collision energy.
The variable $\alpha$ controls the balance between two components: participant
scaling and binary collision scaling. With
the value of $\alpha=0.13$, one may obtain a nice description of the centrality
dependence of average charged particle
multiplicity at midrapidity $|\eta|<0.5$ in Au+Au collisions  at
$\sqrt{s_{NN}}=200$~GeV \cite{Back:2002uc} (see
Fig.~\ref{ncharge_npart}).

\begin{figure}[htb]
\includegraphics[width=8cm]{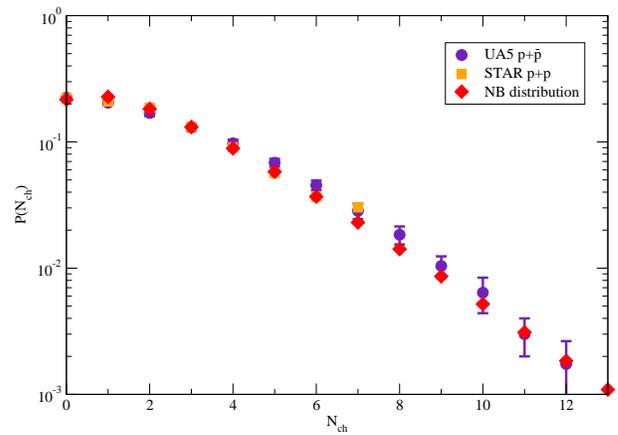}
 \caption{(Color online) Charged particle multiplicity distribution at
midrapidity $|\eta|<0.5$ in p+$\bar{\rm p}$ collisions
 and p+p collisions at $\sqrt{s_{NN}}=200$~GeV.
} \label{ncharge_pp}
\end{figure}

As is well known, the particle multiplicity in high energy collisions is
fluctuating from one event to another. The
distribution of particle multiplicities $N(\Delta \eta)$ for a given rapidity
range $\Delta \eta$ may be well described
by a negative binomial (NB) distribution, \cite{Ansorge:1988kn, Gans, Adare:2008ns},
\begin{align}
P(N, \mu, k) = \frac{\Gamma(N+k)}{\Gamma(N+1) \Gamma(k)}
\frac{(\mu/k)^{N}}{(\mu/k + 1)^{N+k}} \ \ \ \
\label{NBD}
\end{align}
where $\mu$ is the mean of the distribution, and $k$ is related to the shape of
the distribution. The variance is
given by $\sigma^2=\mu(\mu/k+1)$ and the scaled invariance is defined as
$\omega=\sigma^2/\mu=\mu/k+1$.
With the values of $\mu = 2.35$ and $k=1.9$, one obtains a good description of
the charge particle multiplicity
measurements for both p+$\bar{\rm p}$ collisions from UA5 \cite{Ansorge:1988kn},
and p+p collisions from STAR
\cite{Gans} at midrapidity $|\eta|<0.5$ (see Fig.~\ref{ncharge_pp}).

For high energy nucleus-nucleus collisions, we include particle multiplicity
fluctuations by evaluating Eq.(\ref{2comp_nptc})
on an event-by-event basis, with the distribution of $N_{NN}(\Delta \eta)$ given
by Eq.(\ref{NBD}). The balance factor $\alpha$ of
binary collision and participant scaling in Eq.(\ref{2comp_nptc}) is implemented
by randomly keeping only a fraction $\alpha$ of
particles from a binary collision, and a fraction $(1-\alpha)/2$ of particles
originating from a participating nucleon. In general the
positions of particles might be sampled according to a smeared distribution
around the positions of binary collisions or
participant nucleons. Here we take the positions of particles as the same
positions as binary collisions or
participant nucleons.
The Lorentz contraction in the longitudinal direction is taken into account by
contracting the longitudinal position
$z$ of each particle by a factor of $100$ for Au+Au collisions at
$\sqrt{s_{NN}}=200$~GeV.

With the number of produced particles and their positions fixed, we also assign
momenta to each particle. In this
work, particle transverse momenta $p_T$ are sampled according to the following
power law distribution,
\begin{align}
\frac{dN}{dp_T^2} = \frac{a}{(1 + p_T^2/b^2)^c}
\end{align}
where $a$ is the normalization constant, and $b$ and $c$ are taken as $b=0.88$
and $c=4$. The azimuthal angle of the transverse momentum is uniformly
distributed. Particle rapidities are taken to be uniformly distributed around
mid-rapidity $|y| < 1$ (When changing
the rapidity range, we change both the mean $\mu$ and the parameter $k$ and keep
the scaled variance $\omega$ of NB
distribution fixed). Particles with large rapidities will be absent from the
central rapidity region at later times and
we neglect them in this work. With the above setup, we obtain the full phase
distribution of the system at initial production time.

\section{Initial Geometry}

\begin{figure}[htb]
\includegraphics[width=8cm]{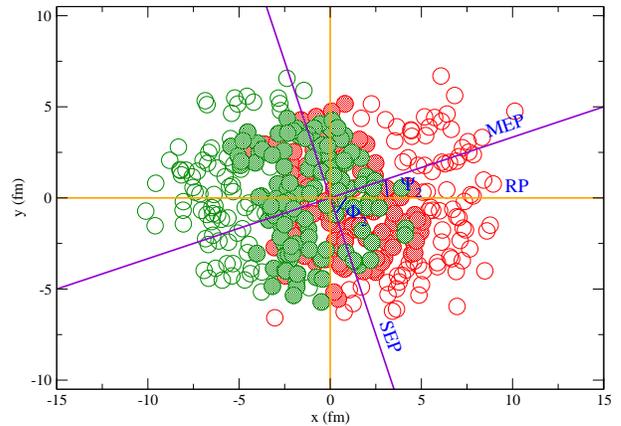}
 \caption{(Color online) The transverse plane for one typical collision event, where the cycles represent nucleons from two nuclei, 
with shaded ones for participating nucleons. Also shown are the locations of different planes: 
the reaction plane (RP), the spatial event plane (SEP) and the momentum event plane (MEP)  for $n=2$. 
} \label{phin_psin}
\end{figure}

Before moving to the evolution of the system, we first
investigate its geometrical properties at the production time.
In a nucleus-nucleus collision, the reaction plane is defined by the beam
direction ($z$) and the impact parameter
direction ($x$). The impact parameter direction and the third orthogonal
direction ($y$) define the transverse plane (one typical collision event is shown Fig. \ref{phin_psin}). 
We call the plane defined by $z$ direction and $y$ direction the vertical plane. 
The geometry of the transverse plane is particularly interesting due to the fact
that the elliptic flow $v_2$ is found in ideal hydrodynamics to be proportional to
the initial eccentricity
$\epsilon_2$ of the overlap region of the colliding nuclei, the determination of
which plays an important
role in the extraction of the transport coefficients of the produced fireball.

For averaged initial conditions, the geometry of the system can be studied
directly in the above framework due to
the coincidence of the vertical plane and the spatial event plane for $\epsilon_2$
(a rotation by $\pi/2$ of the participant plane if the participating nucleons are considered for the spatial distribution).
With fluctuating initial conditions, the spatial event plane is tilted with respect to the reaction plane from one
event to another. We call the angle between the spatial event plane and the
reaction plane the spatial event plane angle $\Phi_2$. 
Note that this choice of the spatial event plane is convenient when generalizing to higher moments
since the event plane angle distribution always has a maximum in the $y$ direction for all even moments 
whereas not always one of the minima is in the $x$ direction.

The final elliptic flow $v_2$ is defined with respect to a third plane, the momentum event plane $\Psi_2$
 that is reconstructed in experiments from the measured momentum distribution of the produced particles. 
Again in ideal hydrodynamics with smooth initial conditions 
this event plane coincides with the reaction plane, i.e., is rotated with respect to the spatial event plane by $\pi/2$. 
This rotation ensures that the final $v_2$ has the same sign as the initial $\epsilon_2$. 
If an event-by-event analysis with fluctuating initial conditions is applied, a strong correlation of the final momentum event
plane to the initial spatial event plane still remains, but fluctuates around $\pi/2$ as has been shown in \cite{Holopainen:2010gz}.

One may generalize the above concept for every harmonic
moment, and define the spatial anisotropy
parameters $\epsilon_n$ as follows. The first moment $\epsilon_1$ can always be
made to vanish by shifting the
coordinates to the center of mass (CM) frame of the  system such that $\langle x
\rangle = \langle y \rangle =0$.
Note $\langle ... \rangle$ throughout this paper represent averages over the
phase space profile for a given event, except in the Appendix.
Once the system is shifted to its CM frame, $\epsilon_1=0$, all higher harmonic
moments can be defined as:
\begin{align}
\epsilon_n = {\sqrt{\langle r_\perp^n \cos(n\phi) \rangle^2 + \langle r_\perp^n
\sin(n\phi) \rangle^2}} / {\langle
r_\perp^n \rangle}
\end{align}
where $r_\perp = \sqrt{x^2+y^2}$, and $\phi=\arctan(y/x)$ are polar coordinates
in the transverse plane.
The spatial event plane angle $\Phi_n$  with respect to the reaction plane can
be found through the following formula,
\begin{align}\label{Phin}
\Phi_n = \frac{1}{n} \arctan \frac{\langle r_\perp^n \sin(n\phi)\rangle}
{\langle r_\perp^n \cos(n\phi) \rangle}
\end{align}
Note in our definition $\Phi_n$ fluctuates from one event to another in the
range of ($-\pi/n, \pi/n$), but it is equivalent to rotate such angle by $2\pi/n$.
Once the event plane angle is found, the definition of the spatial anisotropy
parameters may be reduced to
$\epsilon_n = {\langle r_\perp^n \cos[n(\phi-\Phi_n)] \rangle} / {\langle
r_\perp^n \rangle}$.

The flow coefficients $v_n$ are defined as the $n$-th Fourier moment of the
particle momentum distribution with respect to
each momentum event plane,
\begin{align}
v_n = \langle \cos[n(\psi - \Psi_n)] \rangle
\end{align}
where $\psi = \tan^{-1}(p_y/p_x)$ is the azimuthal angle of particle momentum
$p$ in the CM frame.
Here we define the momentum event plane by a rotation angle $\pi/n$ with respect to the initial spatial event plane, $\Psi_n=\Phi_n+\pi/n$. 
This rotation is just a convention generalized from the requirement that 
a positive initial eccentricity generates a positive value of final elliptic flow in ideal hydrodynamics with average initial conditions.
Note our definition of the momentum event plane does not necessarily corresponds to the event plane that is reconstructed in experiments as just mentioned, but 
for our systematic study it provides an unambiguous basis to quantify the final state response to the initial state anisotropies.

\begin{figure}[htb]
\includegraphics[width=8cm]{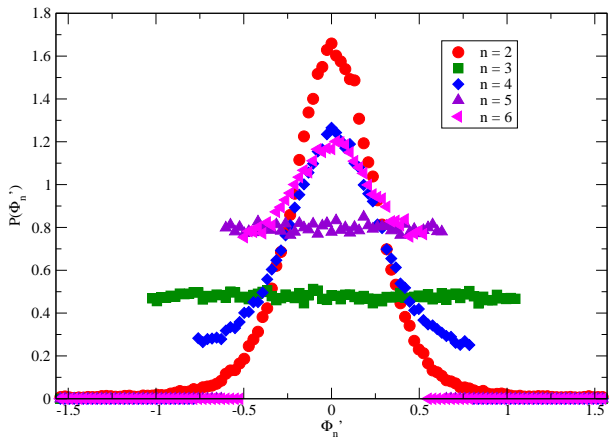}
 \caption{(Color online) The probability distribution of $\Phi_n'$ at production
time with $b=8$~fm.
} \label{ini_phin_ebe}
\end{figure}

Since the event plane defined by Eq. (\ref{Phin}) fluctuates around the vertical plane ($y$ direction) for even moments, 
we may perform a transformation $\Phi_n \to \Phi_n'$ by evaluating Eq. (\ref{Phin}) 
with $\phi \to \phi'=\phi+\pi/2$. This corresponds to a rotation of the 
coordinate system $(x,y)$ by $\pi/2$ which ensures the distributions of 
all even $\Phi_n'$ peak at $0$ as shown in Fig.~\ref{ini_phin_ebe}. 
In this plot, the impact parameter $b$ is taken to be $8$~fm for all events.
While all even moments are strongly correlated with the reaction plane with one maximum along $y$ direction, 
all odd moments are uniformly distributed. 
This may be understood since the odd moments of spatial anisotropy purely originate
from fluctuations while the even ones are combined effects of fluctuations and geometry.
As a consequence, if one defines the spatial anisotropy parameters $\epsilon_n$
with respect to the pre-determined the reaction plane, the event-averaged
$\epsilon_n$ vanishes for all odd moments, but not for even ones.
We also observe that the distributions of even moments is wider for higher
values of $n$ due to weaker correlations with respect to the reaction plane 
[in fact, what matters is the distribution of $n\Phi_n'$, as $\Phi_n'$ fluctuates within $(-\pi/n, \pi/n)$].

\begin{figure}[htb]
\includegraphics[width=8cm]{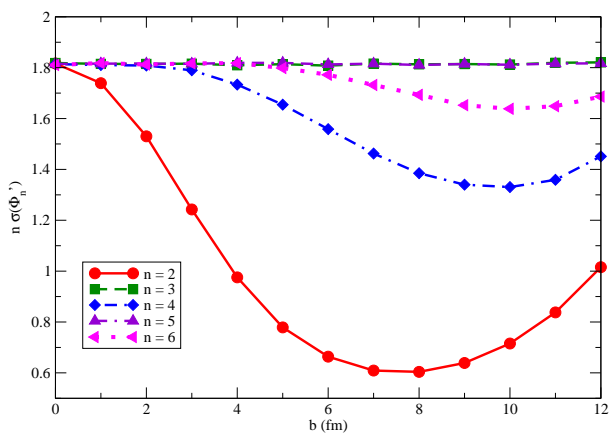}
 \caption{(Color online) The widths of $\Phi_n'$ distribution (times $n$) at
production time as a function of impact parameter $b$.
} \label{ini_phin_sigma}
\end{figure}

We also investigate the centrality dependence of the above correlations in Fig.~\ref{ini_phin_sigma}, where the widths of
the distributions are plotted as a function of impact parameter $b$.
One can see that the widths all odd values of $n$ align with each other at $\pi/\sqrt{3}$
as expected [for a uniform distribution
from $-\pi/n$ to $\pi/n$, the variance is $\sigma^2=\pi^2/(3n^2)$ and we are
plotting $n \sigma(\Phi_n')=\sigma(n\Phi_n')$].
Also due to symmetry in central collisions there is no
correlation between the angle $\Phi_n'$ and $y$ direction for
all values of $n$. The anisotropy is purely from fluctuations,
rendering uniform distributions also for even values of $n$.
As one moves to non-central collisions, geometry comes into play and may
dominate over pure fluctuations,
hence the widths of even $n$ distributions become smaller.
For very peripheral collisions, the importance of the geometry diminishes due to
the small size of the system,
and even $n$ distributions become broader again.
We also observe that even harmonic moments with higher values of $n$ have weaker
dependence on centrality.

\begin{figure}[htb]
\includegraphics[width=8cm]{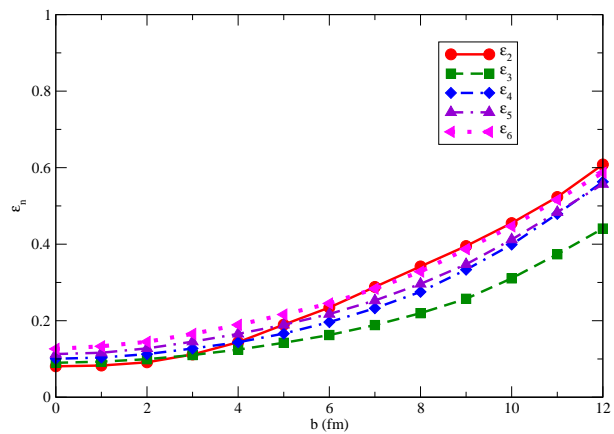}
 \caption{(Color online) The spatial anisotropy parameters $\epsilon_n$ at
production time as a function of centrality from
 Monte Carlo Glauber modelling of Au+Au collisions at $\sqrt{s_{NN}}=200$~GeV (see text for details).
} \label{ini_epsn}
\end{figure}

In Fig.~\ref{ini_epsn}, the first few spatial anisotropy parameters $\epsilon_n$
are plotted as a
function of impact parameter $b$ ($\epsilon_1=0$ is not shown).
One may observe that all moments are of the same magnitude for typical
non-central collisions. In central collisions, as pure
fluctuations instead of geometry generate the anisotropy, higher moments acquire
larger values due to larger fluctuations brought by the power $n$
in the definition of $\epsilon_n$. Note that if the same weight, i.e.,
$r_\perp^2$, is taken for every $\epsilon_n$ as in
Ref. \cite{Alver:2010gr}, all moments are the same in central collisions and
$\epsilon_2$ is larger than all higher
moments in non-central collisions (also note $\epsilon_1$ is non-zero if
$r_\perp^2$ is used).

\begin{figure}[htb]
\includegraphics[width=8cm]{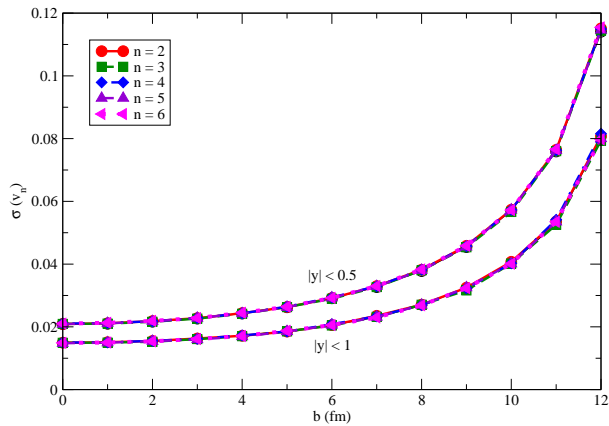}
 \caption{(Color online) The widths of the initial $v_n$ distribution at the
production time as a function of impact parameter $b$.  
} \label{ini_vn}
\end{figure}

In our initial conditions, we may also calculate the momentum anisotropies as we
have generated the full phase space distribution for the produced system.
Since the initial particles are sampled with a symmetric azimuthal distribution,
one obtains zero $v_n$ when averaging over events.
In Fig. \ref{ini_vn}, the width of the initial $v_n$ distribution is plotted as
a function of impact parameter $b$. 
We find that the width increases as one moves from central collisions to
non-central collisions due to the decrease in the number of particles in the
produced system.
In fact, the $v_n$ distribution is a Gaussian as a result of central limit
theorem and its width is found to be $1/\sqrt{2N}$ for all values of $n$, except
for 
very peripheral collisions where the particle number $N$ is too small.
The figure contains two sets of curves: the upper one is for all particles within a
rapidity bin of $|y|<0.5$ and the lower one for $|y|<1$. The curves are related
by a factor of $\sqrt{2}$  since the number of particles in the system is
doubled via the doubling
of the rapidity bin.
The non-zero width of initial $v_n$ distribution helps to explain the wide
distribution of the transformation matrix elements 
between initial $\epsilon_n$ and final $v_n$ as shown in Fig.
\ref{Tvnepsn_dist}. 
It serves as another source that contributes to final flow fluctuations in addition to initial state geometry fluctuations.

\section{Pre-Equilibrium Phase}

As to date, hydrodynamical simulations mostly use initial conditions calculated at
the initial production time of the medium, and have neglected
the influence of the pre-equlibrium time evolution of the colliding matter.
In this sense, the spatial information inferred from comparing experimental
measurements with hydrodynamical simulations
is for the system at the starting time of the hydrodynamic evolution $t=t_0$, not at the initial
production time $t=0$.
However, the pre-equilibrium evolution may be important to include when
considering the geometry fluctuations of the produced matter.
Unlike the hydrodynamical evolution which directly translates the initial
geometric anisotropies into the observed momentum
anisotropies, the early pre-equilibrium expansion of the system will not only
smear out the spatial fluctuations and change the local momentum distribution,
but may also lead to correlations between odd and even moments. The inclusion of
the pre-equilibrium evolution could be also important for studying
the Hanbury-Brown--Twiss interferometri radii as it will generate some amount of early flow 
\cite{Jas:2007rw, Broniowski:2008qk, Broniowski:2008vp, Pratt:2008qv}.

To simulate the pre-equilibrium evolution, we solve the Boltzmann equation for
the phase space distribution
$f\left({\bf x}, {\bf p}, t\right)=dN/d^3{\bf x}d^3{\bf p}$ of the system,
\begin{align}
\left({\partial_t}+ {\bf v}\cdot {\bf \nabla}_{\bf x} \right) f({\bf x}, {\bf
p}, t) = C[f] \ \ \ \ \ \
\end{align}
Here, for simplicity massless particles are considered, $|{\bf v}|=1$.
The free-streaming term  will smear out the spatial fluctuations and change the
local momentum distribution due to pure fluctuations.
The collision term is important for studying the details of the thermalization of the system, 
a complex issue which has not been fully understood yet.
In this work, we focus on the effect of the pre-equilibrium expansion on the
system geometry and only include the free-streaming term by setting $C[f]=0$.
Such a treatment is important  for our study as the flow built up by
hydrodynamical evolution is mostly driven by
spatial anisotropy. The consideration of the collision term will remain for a
future project.
The Boltzmann equation containing only the free-steaming term can be solved
analytically, with the streaming solution given by
$f({\bf x}, {\bf p}, t) = f({\bf x} - {\bf v} (t-t_i), {\bf p}, t_i)$, where
$t_i$ is the initial starting time of the evolution.

\begin{figure}[htb]
\includegraphics[width=8cm]{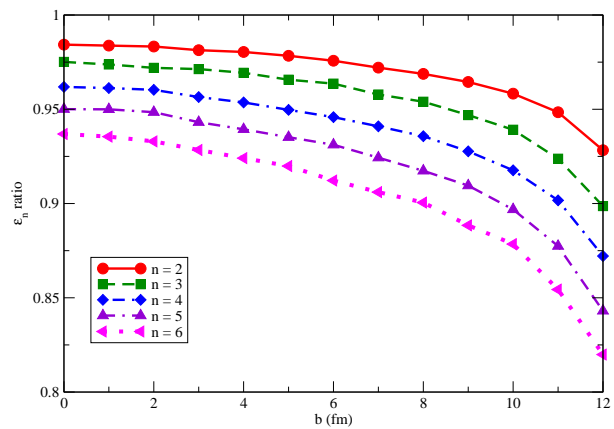}
 \caption{(Color online) The ratio of $\epsilon_n$ at $t_0=0.6$~fm/c to those at
initial production time
 as a function of impact parameter $b$. } \label{preeq_epsn_ratio}
\end{figure}

The effect of free-streaming on the spatial anisotropies during the early
expansion is shown in Fig. \ref{preeq_epsn_ratio}, where the ratios
of anisotropy parameters $\epsilon_n$ evaluated at $t_0=0.6$~fm/c to those at
the production time are shown as a
function of centrality. As expected, the expansion of the matter due to
free-streaming smears out the spatial
fluctuations: all the spatial anisotropy parameters $\epsilon_n$ become smaller.
This diminishing effect is more pronounced
in non-central collisions due to the smaller size of the system. We also observe
that higher moments get more diminished than lower
moments.

\begin{figure}[htb]
\includegraphics[width=8cm]{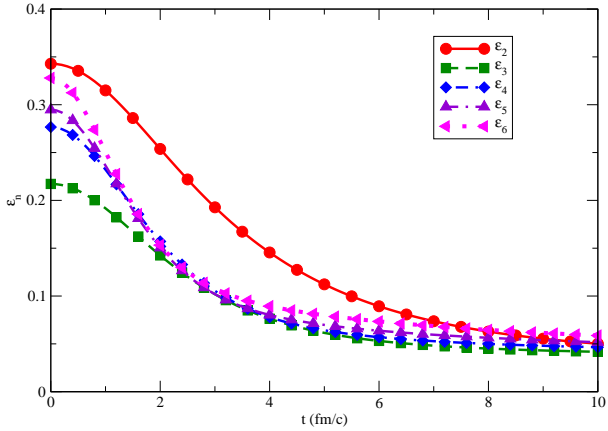}
 \caption{(Color online) Time evolution of $\epsilon_n$ when particles are just
freely streaming for an impact parameter of $b=8$~fm.  } \label{preeq_epsn_avr}
\end{figure}

We further explore the time evolution of the above smearing effect due to
free-streaming as shown in Fig. \ref{preeq_epsn_avr}, where
$\epsilon_n$ at an impact parameter of $b=8$~fm is plotted as a function of
time. We observe that
the anisotropy parameters $\epsilon_n$ decrease rather fast for the first
$2-3$~fm/c and then slowly saturate. The observed reduction of the $\epsilon_n$
hints at the importance  of including the pre-equilibrium expansion when
studying the initial state geometry fluctuations for hydrodynamical
simulations. The relative size of the different coefficients even depends on the 
duration of the pre-equilibrium expansion and the relative size of the resulting 
flow coefficients might be used to constrain this initial evolution. 

As we just mentioned, there are two separate effects due to the pre-equilibrium
evolution: the pure drift effect due to the system expansion
which tends to diminish all moments, and the correlations between odd and event
moments which is absent in the later hydrodynamical evolution.
The mixing effect between odd and event moments can be clearly seen when one
explicitly performs a multipole-expansion analysis for the Boltzmann equation.
To start, we choose  spherical polar coordinates for both coordinate space ${\bf
x} = (r,\theta, \phi)$ and momentum space
${\bf p} = (p, \theta_p, \phi_p)$. To make the analysis dimensionless, we define
$\tilde{r} = r/r_{\rm max}$
and $\tilde{p} = p/T_p$, where $r_{\rm max}$ and $T_p$ are two dimensional
quantities (being constants or varying with time).
Then we expand the phase space distribution as
\begin{align}
\hspace{-24pt} f(\tilde{\bf x},\tilde{\bf p},t) = \sum_{nlmNLM} a_{nlm}^{NLM}(t)
R_{nl}(\alpha_{nl}, \tilde{r})
 Y_{lm}(\theta,\phi)  &  \nonumber\\  \exp({-\tilde{p}})  P_{N} (\tilde{p})
Y_{LM}(\theta_p, \phi_p)&
\end{align}
In the above expression,
\begin{align}
& R_{nl}(\alpha_{nl}, \tilde{r}) = \frac{\sqrt{2}j_l(\alpha_{nl} \tilde{r})
}{j_{l+1}(\alpha_{nl})} \nonumber\\
& P_{N}(\tilde{p}) = \sqrt{\frac{N!}{(N+\mu)!}} L_N^{(\mu)}(\tilde{p}) 
\end{align}
where $j_l$ is the spherical Bessel function with $\alpha_{nl}$ the $n$th root
of function $j_l$, $L_N^{(\mu)}$ the $\mu$-th order
Laguerre function (Here we choose $\mu = 2$) and $Y_l^m(\theta,\phi)$ the
spherical harmonics. The expansion coefficients $a_{nlm}^{NLM}$ are determined
from the phase space distribution by
\begin{eqnarray}
\hspace{-24pt} a_{nlm}^{NLM}(t) = \int_0^{1} \tilde{r}^2 d\tilde{r} \int d\Omega
\int_0^\infty \tilde{p}^2 d\tilde{p}
\int d\Omega_p R_{nl}(\alpha_{nl}, \tilde{r})
&  \nonumber \\
Y^*_{lm}(\Omega) P_{N} (\tilde{p}) Y^*_{LM}(\Omega_p) f(\tilde{\bf x},\tilde{\bf
p},t) \ \ &
\end{eqnarray}
The spatial anisotropy parameters $\epsilon_m$ are related to the expansion
coefficients $a_{nlm}^{NLM}$, by
\begin{eqnarray}
 \langle (\sin\theta)^m \sin(m\phi) \rangle \!\!&=\!\!&  -\frac{ C[m]}{N}
\sum_{n} {\rm Im} \left[a^{000}_{nmm}\right] J_r[n, m] \ \ \ \ \ \ 
\nonumber\\
\langle (\sin\theta)^m \cos(m\phi) \rangle \!\!&=\!\!&  \frac{ C[m]}{N} 
\sum_{n} {\rm Re}
\left[a^{000}_{nmm}\right]  J_r[n,m] 
\end{eqnarray}
where
\begin{align}
&C[m] = (-1)^{m} \sqrt{\frac{4\pi (2m)!}{2m +1}} \frac{\sqrt{8\pi}}{(2m-1)!!}
\nonumber\\
&J_r[n, m] = \frac{\sqrt{2}}{j_{m+1}(\alpha _{{nm}})} \int _0^1 \tilde{r}^2
d\tilde{r} j_{m} (\alpha_{nm} \tilde{r})
\end{align}
The normalization factor $N$ represents the total number of particles in the
system, which is given by
\begin{align}
N = \sqrt{4\pi} \sqrt{8\pi} \sum_{n} a^{000}_{n00}  J_r[n,0]
\end{align}
Note for $\langle (r \sin\theta)^m \cos(m\phi) \rangle$ and $\langle
(r\sin\theta)^m \sin(m\phi) \rangle$ which appear in the
definition of $\epsilon_m$, there will be an extra factor $r^m$ in the integral
$J_r$.

Within the above multipole expansion analysis, the Boltzmann equation becomes,
\begin{eqnarray}
 \frac{\partial a^{N'L'M'}_{n'l'm'}(t)}{\partial t}
+ \sum _{nlmNLM} a^{NLM}_{nlm}(t) \frac{\alpha_{nl}}{r_{\rm max }}
I_r[n',l',n,l]   & \nonumber\\
 \delta_{NN'} \left(\delta_{l,l'-1} + \delta_{l,l'+1} \right)
\left(\delta_{L,L'-1}+\delta _{L,L'+1}\right) & \nonumber
\\    \sqrt{\frac{l+l'+1}{2(2l+1)}} \sqrt{\frac{2L+1}{2L'+1}} (L,0;1,0|L',0) &
\nonumber\\ \!\sum_{i}\! \delta_{m,m'+i
}  \delta_{M,M'-i} (l',m';1,i |l,m) (L,M;1,i |L',M') & \nonumber \\ =
C[a^{N'L'M'}_{n'l'm'}]  \ \ \ &
\end{eqnarray}
where
\begin{align}\label{Ir_nlnl}
I_r[n',l',n,l] =  \frac{\sqrt{2}}{j_{l'+1}(\alpha _{n'l'})}
\frac{\sqrt{2}}{j_{l+1}(\alpha _{{nl}})} & \nonumber\\
\int _0^1 \tilde{r}^2 d\tilde{r} j_{l'} (\alpha_{n'l'} \tilde{r}) j_{l'}(\alpha
_{{nl}} \tilde{r}) &
\end{align}
Note the indices in $j_{l'}(\alpha _{{nl}} \tilde{r})$ in
$I_r^{(0)}[n',l',n,l]$, which do not allow us to perform the
integral using orthogonal relations. $(l_1,m_1;l_2,m_2|j,m)$ are the
Clebsch-Gordan coefficients for adding two angular
momenta ${\bf j} = {\bf l}_1 + {\bf l}_2$. More details of the derivation are
presented in the Appendix.

\begin{figure}[htb]
\includegraphics[width=8cm]{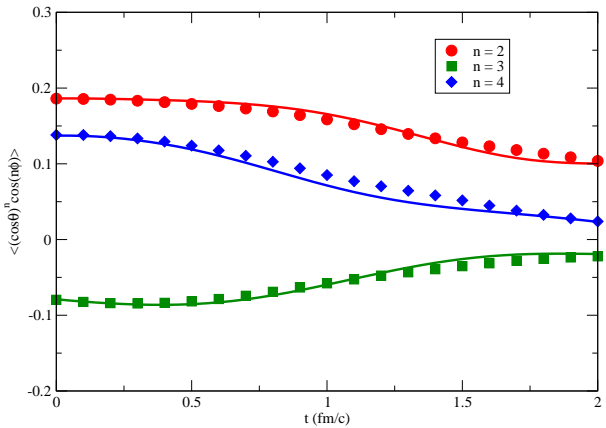}
 \caption{(Color online) Time evolution of $\langle (\sin\theta)^n
\cos(n\phi)\rangle $ from multipole expansion
 analysis (lines) and from directly solving the free-streaming term of Boltzmann
equation (symbols) for an event with $b=8$~fm.
} \label{preeq_epsn_multipole}
\end{figure}

From the above equations, one immediately sees the mixing between odd and even
moments both for spatial part ($l \to l\pm 1$)
and momentum part ($L \to L\pm 1$) due to the free-streaming of particles. We
also check the above expression by comparing numerically with the result from
directly solving free-streaming part of Boltzmann equation. This is shown in
Fig. \ref{preeq_epsn_multipole}, where we
plot the time evolution of $\langle (\sin\theta)^n \cos(n\phi) \rangle$ for one
typical event with impact parameter $b=8$~fm and see that the two results
nicely agree with each other.

In order to further separate the correlation effect from the pure drifting
effect during the pre-equilibrium expansion, we
perform the following analysis. We relate the spatial anisotropies at two
different times by a transformation matrix,
\begin{align}
\left(%
\begin{array}{c}
  \epsilon_2(t_0) \\
  \epsilon_3(t_0) \\
\end{array}%
\right) =
\left(%
\begin{array}{cc}
  D_{22}(t_0) & D_{23}(t_0) \\
  D_{32}(t_0) & D_{33}(t_0) \\
\end{array}%
\right) \left(%
\begin{array}{c}
  \epsilon_2(0) \\
  \epsilon_3(0) \\
\end{array}%
\right)
\end{align}
Here we only consider the second and third moments --  the inclusion of higher
order moments is straightforward and is expected to give only
small contributions which we neglected in the current analysis. 
The diagonal elements of the transformation matrix quantify the pure drifting
effect and the off-diagonal elements represent the effect of
the mixing between the second and third moments. We obtain the distribution of
the transformation matrix elements by
pairing two linear independent events from a large set of events.

\begin{figure}[htb]
\includegraphics[width=8cm]{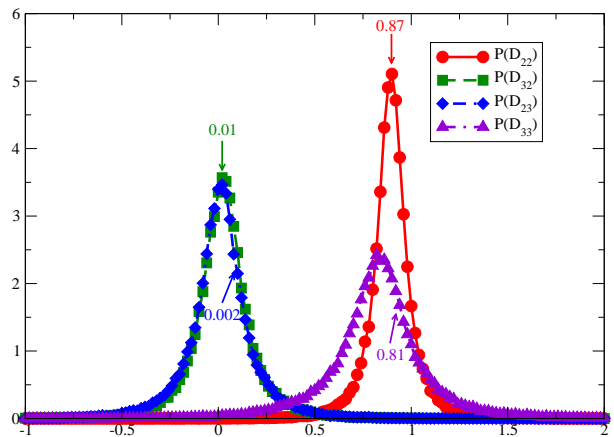}
 \caption{(Color online) The probability distributions of the transformation
matrix between $\epsilon_2$ and $\epsilon_3$
 with the early expansion time taken as $t_0=1.2$~fm. The numbers represent the
mean of each distribution. }
\label{Teps_dist}
\end{figure}

\begin{figure}[htb]
\includegraphics[width=8cm]{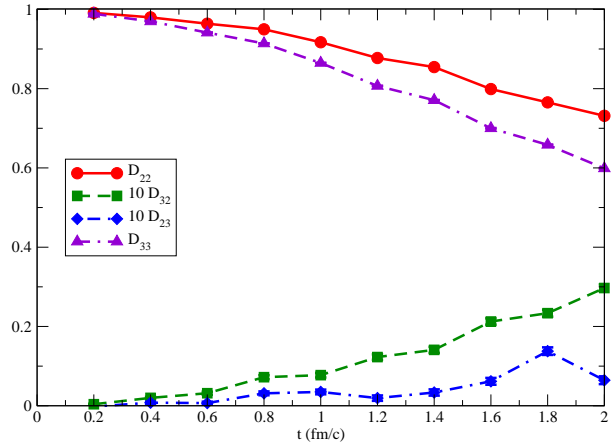}
 \caption{(Color online) Time evolution of the transformation matrix between
$\epsilon_2$ and $\epsilon_3$. }
\label{Teps_evolve}
\end{figure}

In Fig.~\ref{Teps_dist}, we show the probability distribution of four elements
of the above transformation matrix,
with the pre-equilibrium expansion time taken to be $t_0=1.2$~fm/c. The numbers
in the figure represent the mean of each distribution (to which the arrows
point).
The impact parameter is taken to be $8$~fm for all events in this plot.
One clearly observes the smearing effect from the free streaming when one looks
at the distributions of the two diagonal elements.
The pure drifting effect is more pronounced for the third anisotropy parameter
$\epsilon_3$ ($19\%$) than for the second one $\epsilon_2$ ($13\%$),
consistent with the above results.
The two off-diagonal elements are close to zero implying weak correlations
between $\epsilon_2$ and $\epsilon_3$ originating from the
free-streaming of the system. We further investigate the time evolution of these
matrix elements up to $2$~fm/c in Fig.~\ref{Teps_evolve}.
Both the drifting effect and the mixing of even and odd moments tend to increase
with time as the system expands.

\section{Hydrodynamical Evolution}

In the previous sections, we have presented the initial conditions of the system
at production time and simulated the
pre-equilibrium evolution by utilizing the free-streaming approximation. As we
have not included interaction among the produced
particles, the system is still highly non-thermal. Up to know, little knowledge
has been attained about the details of
the thermalization mechanisms in relativistic heavy-ion collisions.
In this work, we follow the common practice to assume a sudden thermalization of
the system at $t=t_0$ and start the hydrodynamical evolution with the initial
conditions obtained above (including the free
streaming evolution from $t=0$ to $t=t_0$). We first calculate the
energy-momentum tensor from the full phase space
distribution $f({\bf x}, {\bf p}, t)$,
\begin{align}
T^{\mu \nu }(x) = \int \frac{d^3p}{E} p^{\mu }p^{\nu } f\left({\bf x}, {\bf p},
t\right)
\end{align}
For our discretized phase space distribution $f({\bf x}, {\bf p}, t) = \sum_{i}
\delta({\bf x} - {\bf x}_i)
\delta({\bf p} - {\bf p}_i)$, the momentum integration $\int d^3 p$ turns into
sums over all particles.
The discretized spatial part is smeared with a Gaussian function in order to
ensure
a sufficiently continuous distribution necessary for the hydrodynamic
simulation,
\begin{eqnarray}
\hspace{-24pt} \delta({\bf x} - {\bf x}_i) \rightarrow \frac{
\exp\left[-\frac{(x-x_i)^2+(y-y_i)^2}{2\sigma_{xy}^2}\right]
}{{2\pi\sigma_{xy}^2}}
 \frac{\exp\left[-\frac{(z-z_i)^2}{2\sigma_{z}^2}\right]}{\sqrt{2\pi\sigma_{z}^2
}}
\end{eqnarray}
where the widths $\sigma_{xy}$ and $\sigma_z$ characterize the granularity of
the system in the transverse and
longitudinal directions. Physically, this procedure can be interpreted as
thermal smearing of the system which should have occurred prior to
thermalization at $t=t_0$. In general the choice of these width parameters
depends on the duration of the pre-equilibrium phase and the thermalization time
at which one starts the hydrodynamic evolution. Different choices of the
smearing width will affect the local density of the system, and
thus influence the spatial anisotropy parameters.

\begin{figure}[htb]
\includegraphics[width=8cm]{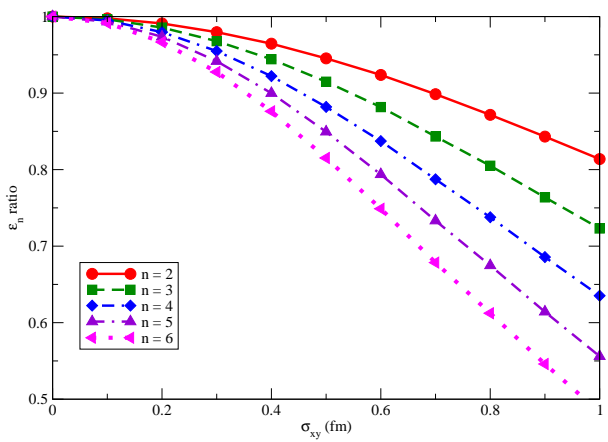}
 \caption{(Color online) The ratio of $\epsilon_n$ with smearing to those
without smearing as a function of the transverse Gaussian width.
} \label{preeq_epsn_smear}
\end{figure}

The effect of the Gaussian smearing on the spatial anisotropy is shown in Fig
\ref{preeq_epsn_smear}, where the ratio of anisotropy parameters
$\epsilon_n$ with smearing to these without smearing is shown as a function of
transverse smearing width
$\sigma_{xy}$. As we are studying the spatial anisotropy in the transverse
plane, the smearing of the longitudinal
direction should be irrelevant and we fix it to be $\sigma_z = 0.5$~fm for our
study.
The impact parameter is taken to be $8$~fm for all calculations shown in this
figure.
As expected, the spatial anisotropies are reduced as one increases the width of
the transverse Gaussian function.
Similar to pre-equilibrium evolution shown before, such smearing effect is
more prominent for higher moments than for lower moments.
Combining both effects (pre-equilibrium evolution and Gaussian smearing), for
typical non-central collisions $\epsilon_2$ may be reduced by about
$10\%$ for a Gaussian width of $\sigma_{xy} = 0.5$~fm and a typical
pre-equilibrium evolution time of $t_0=0.6$~fm; a factor of $2$ larger effect is
observed for for $\epsilon_4$.

In the above construction of the energy-momentum tensor, the initial conditions
at production time are fitted to the final state particle
multiplicity distribution at midrapidity (see Fig. \ref{ncharge_npart} and \ref{ncharge_pp}). Therefore, the energy density of the
system at the thermalization time $t_0$ is underestimated,
due to the longitudinal (and transverse) expansion during the hydrodynamical
evolution.
This effect can be estimated to be about a factor of $2.2$ by directly comparing
our calculation to the final average charged particle multiplicity
$dN_{\rm ch}/d\eta \approx 700$
in central Au+Au collisions at $\sqrt{s_{NN}}=200$~GeV. We have not tuned
our parameters to match the final state particle spectra as we are here not
aiming at 
providing a comprehensive quantitative description of the time-evolution of a
heavy-ion collision, but rather at a targeted study of initial state
fluctuations
and how these initial spatial anisotropies propagate through the fireball
history and translate themselves into collective flow in the final state.

\begin{figure}[htb]
\includegraphics[width=4.2cm]{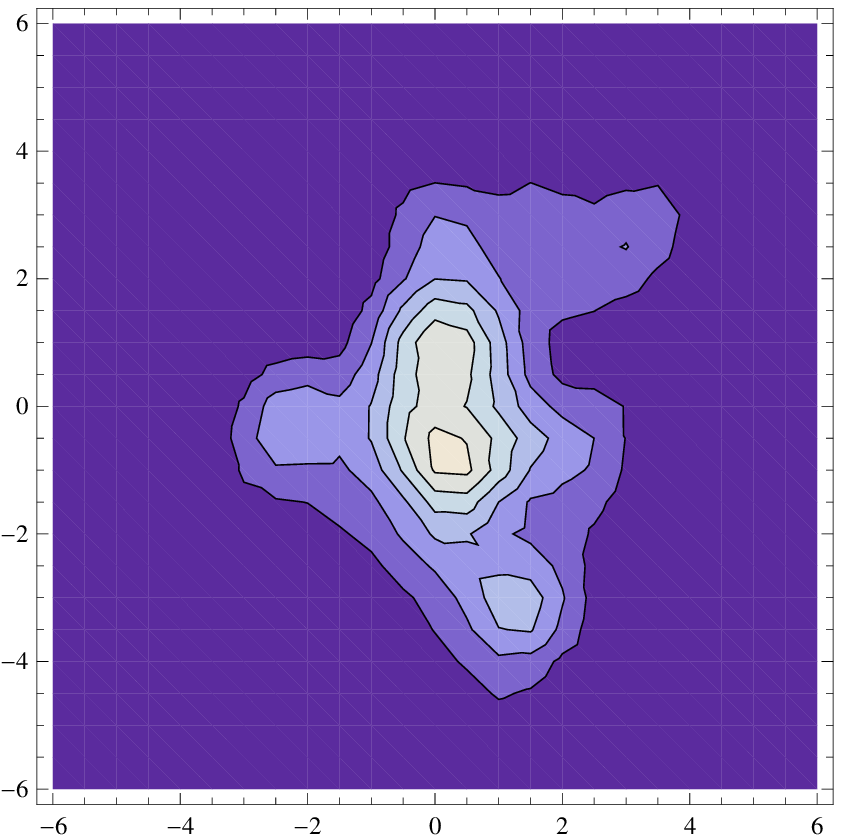}
\includegraphics[width=4.2cm]{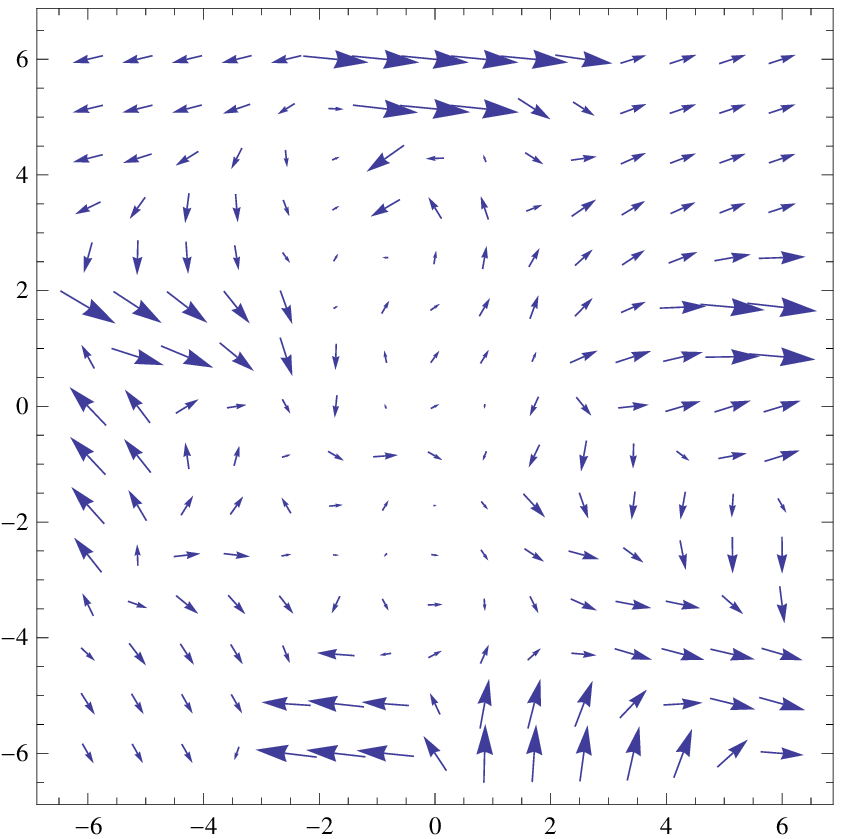}
\includegraphics[width=4.2cm]{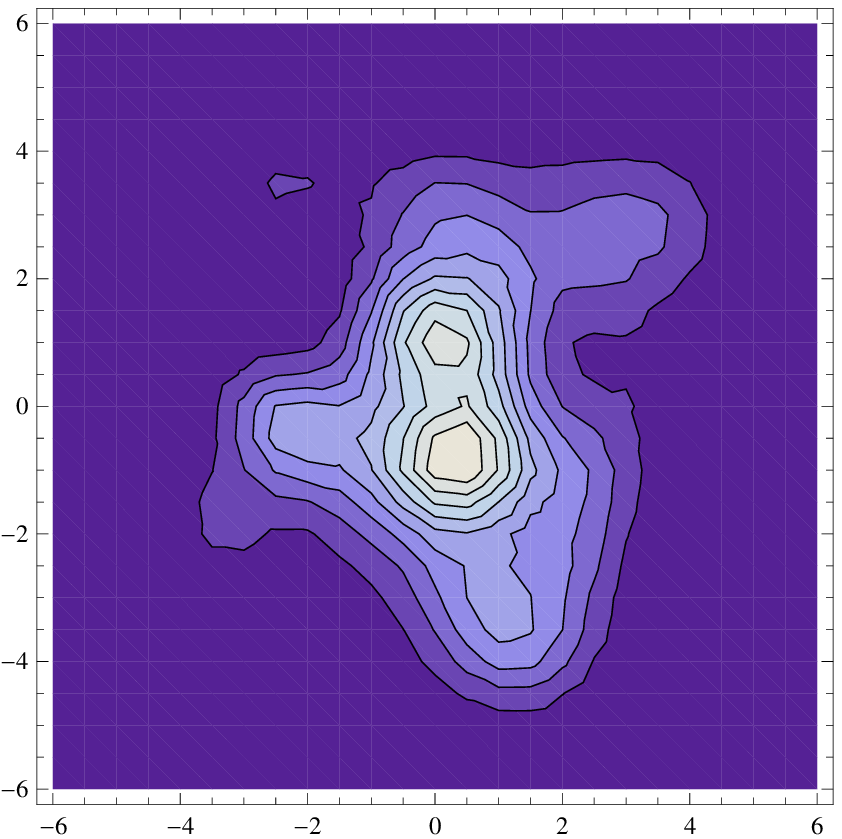}
\includegraphics[width=4.2cm]{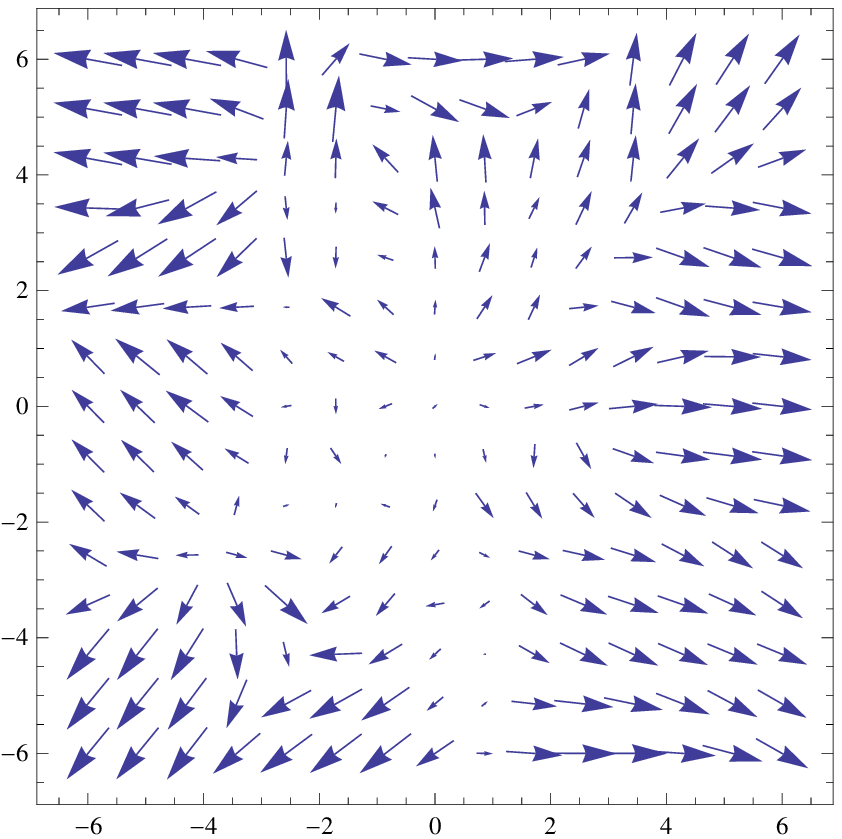}
\includegraphics[width=4.2cm]{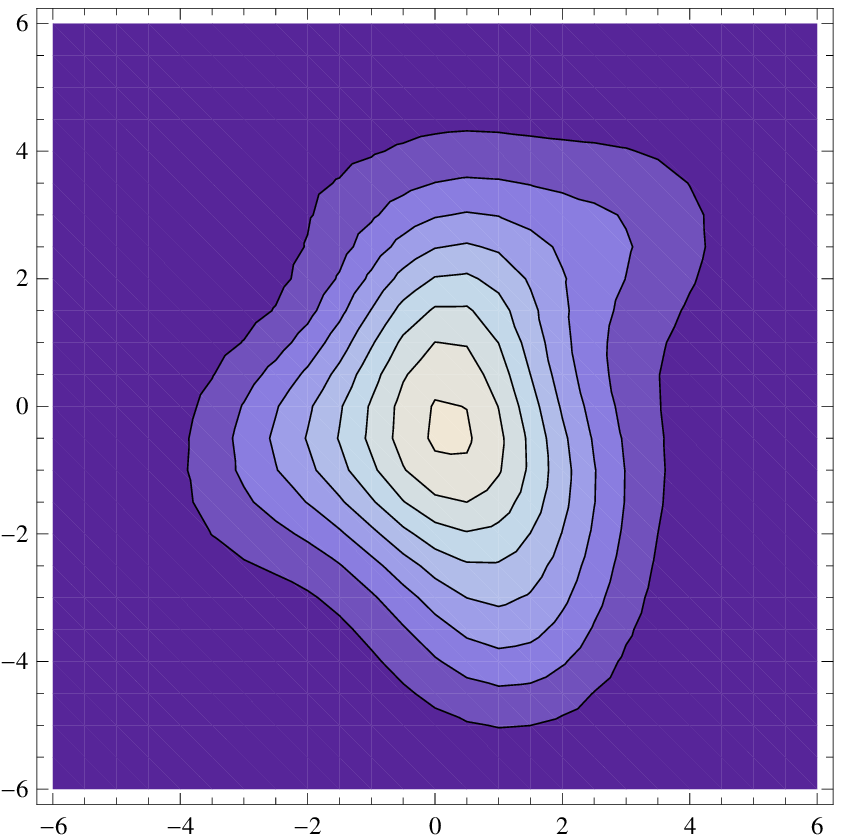}
\includegraphics[width=4.2cm]{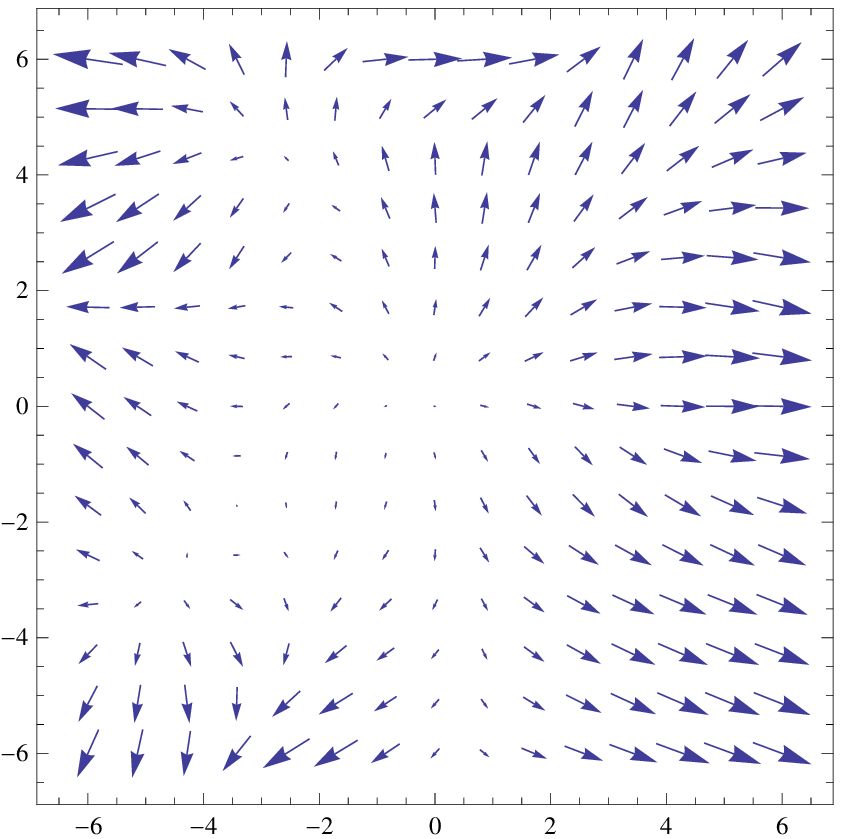}
 \caption{(Color online) The distributions of the energy momentum tensor
components in the transverse plane ($x$, $y$) 
for one typical event with impact parameter $b=8$~fm:   
the left for $T^{00}$ and the right for the flow
vector $(T^{0x}/T^{00}, T^{0y}/T^{00})$. 
Three different sets of parameters are used: 
$t_0=0$, $\sigma_{xy}=0.5$~fm (upper), $t_0=0.6$~fm/c, $\sigma_{xy}=0.5$~fm (middle), 
and $t_0=0.6$~fm/c, $\sigma_{xy}=1$~fm (lower). 
} \label{snapshot_eps}
\end{figure}

In Fig. \ref{snapshot_eps}, we show the a few snapshots of the energy momentum
tensor components in the transverse plane (the horizontal and vertical axes are $x$ and $y$ axes in unit of fm) for one typical event with an impact
parameter $b=8$~fm. 
To illustrate the effects of the pre-equilibrium evolution and Gaussian smearing
of discretized space distribution, we plot three different sets of pre-equilibrium
time and Gaussian smearing width: 
the upper for $t_0=0$ and $\sigma_{xy}=0.5$~fm, the middle for
$t_0=0.6$~fm/c and $\sigma_{xy}=0.5$~fm, 
and the lower for $t_0=0.6$~fm/c and $\sigma_{xy}=1$~fm. 
On the left we show the distribution of $T^{00}$ and on the right the
flow vector $(T^{0x}/T^{00}, T^{0y}/T^{00})$, 
with the arrows representing the directions and the lengths of arrows for the
relative magnitudes of the vectors (within each plot). 
Comparing the left and middle panels, one can clearly see that pre-equilibrium
evolution makes the system larger (thus the energy density becomes smaller) and 
generates some amount of radial flow. The effect of Gaussian smearing
can be seen by comparing the middle and right panels: 
both the energy density and the flow velocity smoothen out significantly when
one increases the Gaussian width.  

After obtaining the energy momentum tensor as described above, we directly start
the hydrodynamical evolution
with the assumption of a sudden thermalization,
\begin{align}
\partial_\mu T^{\mu\nu}(x) = 0
\end{align}
Here an ideal hydrodynamical evolution code \cite{Rischke:1995ir,
Rischke:1995mt} is utilized for our study with a lattice equation of state
\cite{Steinheimer:2009nn, Steinheimer:2009hd} for the hot and dense matter
created in Au+Au collision at $\sqrt{s_{NN}}=200$~GeV.
Particle production at the end of the hydrodynamic evolution 
when the matter is diluted in
the late stage is treated as a gradual freeze-out on an approximated
iso-eigentime hyper-surface according to the Cooper-Frye prescription \cite{Li:2008qm, Steinheimer:2009nn}.
For simplicity, we have not taken into account the hadronic rescattering in the dilute hadron
gas and the resonance decays, since they should not have much influence on the results for 
the charged particle flow coefficients as has been shown in \cite{Petersen:2010md}.

\section{From Initial Geometry Fluctuations to Final Flow}

The above event-by-event setup of the system evolution from initial production
time to the freeze-out of the final state should include all
ingredients that are necessary for the study of the build-up of collective flow 
during the hydrodynamical evolution. 
For the following results, we use the produced charged particles with
transverse momenta $p_T<2$~GeV/c and peudorapidity $|\eta|<1$.

\begin{figure}[htb]
\includegraphics[width=8cm]{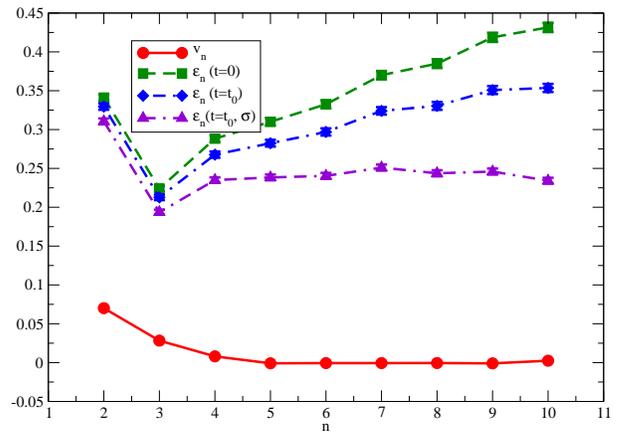}
 \caption{(Color online) The first few spatial anisotropiy parameters
$\epsilon_n$ and flow coefficients $v_n$ as a function of $n$ for $b=5-10$~fm.
 }\label{epsn_vn_vs_n}
 \end{figure}

In Fig. \ref{epsn_vn_vs_n}, we show the first few flow coefficients $v_n$ for
final state particles together with three different spatial anisotropies
evaluated at the production time, at $t=t_0$ before and after Gaussian smearing.
In this figure, the impact parameter is randomly sampled in the $5-10$~fm bin
according to the probability distribution $P(b) \propto b$, the
pre-equilibrium evolution time is set as $t_0=0.6$~fm/c  prior to the
hydrodynamical evolution, and the Gaussian widths for smearing the discretized
initial conditions are taken to be $\sigma_{xy} = \sigma_z = 0.5$~fm. 
One can see that all flow coefficients $v_n$ with $n$ greater than $5$ are
negligible and only the first few $v_n$ ($n=2$, 3, 4)
survive after the hydrodynamical evolution.
We may conclude that the analysis of flow coefficients $v_n$ allows only for
the extraction of the first few spatial anisotropy parameters $\epsilon_n$, but
may not
provide sufficient information to recover the full initial geometry in terms of
all of its higher order harmonics. In order to achieve this, one needs
additional observables with an increased sensitivity to the higher order spatial
anisotropy parameters.
We also note that the spatial asymmetry obtained from comparing flow
measurements to a hydrodynamical simulation only applies to the spatial
characteristics of matter at
the starting time of hydrodynamical evolution $t_0$. In order to obtain the
geometry at the initial production time $t=0$,
one needs to  account for the dynamics of the pre-equilibrium phase -- in our
analysis this would be the smearing effects during the initial free streaming of
the particles and the smoothing of the discretized
initial conditions as shown in the figure.

\begin{figure}[htb]
\includegraphics[width=8cm]{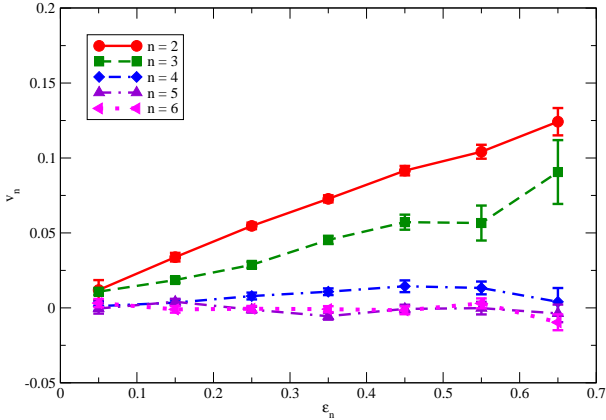}
 \caption{(Color online) The first few flow coefficients $v_n$ as a function of
the same order spatial anisotropy parameter $\epsilon_n$ for $b=5-10$~fm.
 }\label{vn_vs_epsn}
 \end{figure}

In order to study the response of flow build-up to the initial geometry,  Fig.
\ref{vn_vs_epsn} shows the first few flow coefficients $v_n$
as a function of the corresponding spatial anisotropy parameter $\epsilon_n$
evaluated at the production time.
As expected, the hydrodynamical evolution translates spatial anisotropies into
momentum anisotropies, resulting in an essentially linear
relation between $v_n$ and $\epsilon_n$.
We also observe that the curves have smaller slopes for higher moments and
become more or less flat when $n$ is greater than $4-5$. 
This implies that higher flow coefficients show weaker response to the
corresponding geometrical harmonic moments due to larger diminishing effect
originating from the combination of pre-equilibrium evolution, Gaussian
smearing of the discretized spatial distribution,
and the hydrodynamical evolution, all of which tend to lead to a larger
suppression for the higher moments.

\begin{figure}[htb]
\includegraphics[width=8cm]{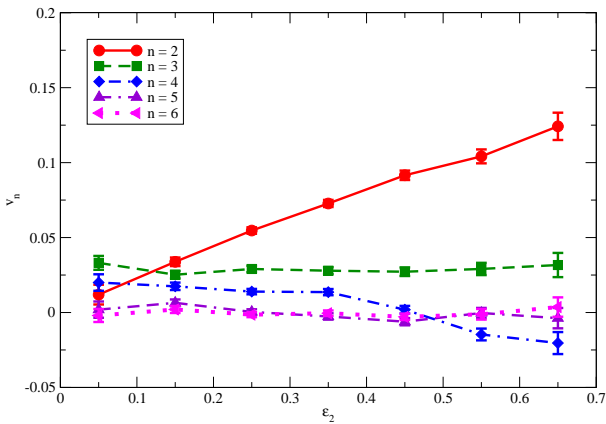}
 \caption{(Color online) The first few flow coefficients $v_n$ as a function of
the second order spatial anisotropy parameter $\epsilon_2$ for $b=5-10$~fm.
} \label{vn_vs_eps2}
\end{figure}

We also explore the correlations between different harmonics moments. As an
illustration, we plot the first few flow
coefficients $v_n$ as a function of the second spatial anisotropy parameter
$\epsilon_2$. On can see that
except for $v_2$ versus $\epsilon_2$, all other curves are essentially flat due
to the combined effect of the small correlations
between odd and even moments and the reduced effect on higher moments from the pre-equilibrium and 
hydrodynamical evolution.

\begin{figure}[htb]
\includegraphics[width=8cm]{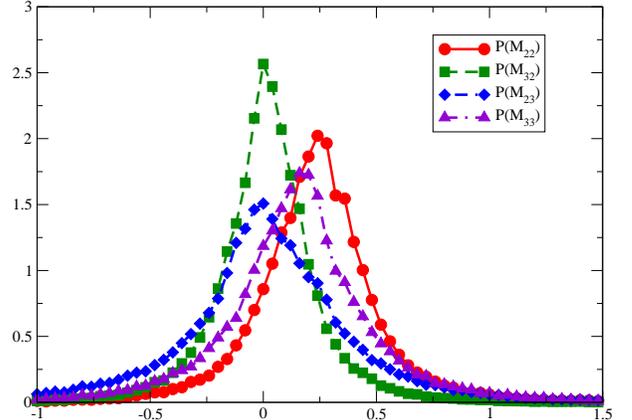}
 \caption{(Color online) The probablity distributions of the transformation
matrix between
 the initial spatial anisotropy parameter $\epsilon_{2,3}$ and the final flow
coefficients $v_{2,3}$
 } \label{Tvnepsn_dist}
\end{figure}

In order to separate the pure smearing effect from the mixing effect between
different moments, we perform an
analysis  similar to that for the pre-equilibrium evolution. On may define the
transformation matrix $M_{nm}$ between the initial spatial anisotropy and the
final
flow coefficients as
\begin{align}
v_{n} = \sum_{m} M_{nm} \epsilon_{m}
\end{align}
where $M_{nm}$ characterizes the strength of the coupling between initial
$\epsilon_{m}$ and final $v_{n}$.
The diagonal elements of the transformation matrix quantify the response of
$v_n$ to $\epsilon_n$ and the off-diagonal elements
represent the effect of the mixing response between different moments $v_n$ and
$\epsilon_{m}$.
Here again we only consider the second and third moment,
\begin{align}
\left(%
\begin{array}{c}  v_2 \\   v_3 \\ \end{array}%
\right) = \left(%
\begin{array}{cc}  M_{22} & M_{23} \\   M_{32} & M_{33} \\ \end{array}%
\right) \left(%
\begin{array}{c}   \epsilon_2 \\   \epsilon_3 \\ \end{array}%
\right)
\end{align}
The extension of this ansatz to include higher order moments is straightforward
and expected to only give small contributions to the dominant moments $n=2,3$.
The distribution of the transformation matrix elements is obtained by solving
two linear independent equations which correspond to a pair of linear
independent events chosen from a large set of events.

In Fig.~\ref{Tvnepsn_dist}, we show the probability distribution of the four
elements of the transformation matrix between final $v_2$, $v_3$ and
initial $\epsilon_2$, $\epsilon_3$.
We find that for the two diagonal elements $\langle M_{22} \rangle_{\rm evt} =
0.21$ is larger than $\langle M_{33} \rangle_{\rm evt} = 0.13$, 
implying stronger response of $v_2$ to $\epsilon_2$ than
$v_3$ to $\epsilon_3$ as expected from Fig. \ref{vn_vs_epsn}. 
The two off-diagonal elements $\langle M_{32} \rangle_{\rm evt}$ and $\langle M_{23}
\rangle_{\rm evt}$ again are very small, implying a weak response of $v_3(v_2)$
to $\epsilon_2(\epsilon_3)$ during the hydrodynamical evolution. 
Another interesting feature is the wide distribution of the transformation matrix which encodes the fluctuations of final flow coefficients. 
We note two initial state effects that contribute to such wide distribution: initial geometry fluctuations and initial $v_n$ fluctuations 
(see Fig. \ref{ini_vn}). 
If one wants to extract information about the initial collision geometry from the measured flow anisotropies, 
it is important to separate the two sources of fluctuations in the transformation matrix. 
Experimentally, this could be achieved by measuring the rapidity correlations of the final flow coefficients
since the initial state geometry fluctuations are expected to be long-range in rapidity, 
while initial state flow fluctuations should decrease when one makes the rapidity window wider.

\begin{figure}[htb]
\includegraphics[width=8cm]{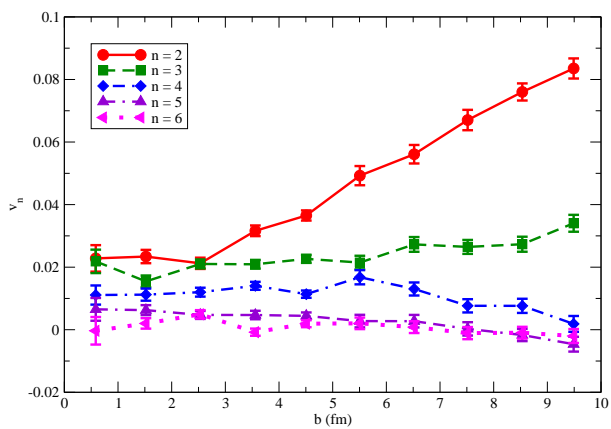}
 \caption{(Color online) The final flow coefficients $v_n$ as a function of
impact parameter $b$.
} \label{vn_vs_b}
\end{figure}

Finally, we explore the centrality dependence of the final state flow
coefficients as shown in Fig. \ref{vn_vs_b}.
The splitting of $v_n$ is clearly seen for all centralities: the lower $v_n$ are
larger than higher $v_n$, 
at least for the first few $v_n$ (when $n\ge 5$, $v_n$ are so small that it
is difficult to resolve their splitting). 
For the most central collisions, our statistics is not sufficient to distinguish
between different curves, but one would expect the same ordering, even
though the splitting might be smaller. This is due to  pure fluctuations being
the only source of spatial anisotropies 
for both odd and event moments in central collisions, and the smearing effect in
the subsequent pre-equilibrium evolution is more prominent for higher moments,
thus
leading to less flow builtup  for higher order $v_n$ during the hydrodynamical
evolution. 
One observes different centrality dependencies for different $v_n$: the lowest
$v_n$ have the strongest centrality dependence.

\section{Summary}

We have presented a systematic study of initial collision geometry fluctuations
and have investigated
how they evolve throughout the whole history of the collision and finally
translate into measurable momentum anisotropies
of the produced hadrons. Our initial conditions at production time $t=0$ are
obtained via a Monte Carlo Glauber model
with the inclusion of nucleon position fluctuations, plus additional
fluctuations stemming  from individual
nucleon-nucleon collisions. In addition we make an ansatz for the initial
transverse momentum distribution of the produced particles which is important
for the treatment
of the pre-equilibrium phase of the collision.
We evolve our full phase space distribution using a Boltzmann equation and
approximate the pre-equilibrium evolution by treating
all particles as free streaming. A sudden thermalization of our initial
conditions is enforced for the subsequent hydrodynamic evolution of 
the thermalized system, which is performed using three-dimensional relativistic
ideal hydrodynamics.

Our analysis shows that though all initial spatial anisotropy parameters are of
the same magnitude,
only the first few flow coefficients for the momentum anisotropy of final state
hadrons actually survive after hydrodynamical
evolution. We also quantitatively investigate the mixing between odd and even
harmonic moments during the pre-equilibrium evolution
and its effect on the evolution of the system asymmetry is found to be small.
The anisotropy of the matter is found to be affected by the pre-equilibrium
evolution and by the smoothing of the
discretized initial conditions necessary for the hydrodynamical evolution (which
can be seen as equivalent to thermal smearing expected to occur during
thermalization), both of which tend to smear out the spatial anisotropies.
The hydrodynamical evolution leads to an additional dampening of the flow
response to the initial spatial anisotropies, particularly for the higher order
moments. This makes it difficult to recover the 
full geometry of the initial state from measuring high order flow coefficients 
which have the ability to provide additional information on the transport
properties of the produced matter. 
We also observe the contribution of initial state flow fluctuations to final flow fluctuations, 
which could be separated by rapidity correlation measurements for a better understanding of initial state geometry fluctuations.

In summary, we have conducted an event-by-event study of the time evolution of
the multipole moments of the collision geometry in a relativistic heavy-ion
collision. 
Our study sheds light on how these multipole moments relate to measurable
collective flow coefficients of the hadronic final state and how the collision
dynamics, both in the pre-equilibrium and in the hydrodynamic evolution phase,
affect the correlation between the initial spatial anisotropies and the final 
momentum space anisotropies. It allows for
improved constraints on the determination of  various transport properties of
the QCD medium, which commonly are extracted by 
analyzing the observed momentum anisotropy of the final particles and are very
sensitive to the proper description of initial spatial anisotropies.
Our current work can be improved in many directions. Here, we have only focused
on the qualitative study of the propagation of the initial state geometry
fluctuations; 
a more quantitative study including a comparison with experimental measurements
would be desirable. 
The pre-equilibrium phase has been approximated by free-streaming of particles;
we fully expect that the inclusion of a realistic collision term in the
Boltzmann equation would provide more sophisticated 
initial conditions for the hydrodynamic evolution. 
We have performed our calculations using ideal hydrodynamics; improving these
with the use of viscous hydrodynamics should help to separate viscosity
dominated effects
from non-viscous effects on the evolution of the geometry fluctuations. All
these tasks we leave to future investigations.

\section{Acknowledgments}

We thank Dirk Rischke for providing the three-dimensional relativistic hydrodynamics code. This work
was supported in part by U.S. department of Energy grant
DE-FG02-05ER41367. Some of the calculations were performed using resources provided
by the Open Science Grid, which is supported by the National Science Foundation and the
U.S. Department of Energy. H.P. acknowledges a Feodor Lynen fellowship of
the Alexander von Humboldt foundation.

\appendix

\section{Multipole Expansion}

In this appendix, we present more details of the multipole expansion of the left
hand side of the Boltzmann equation. The
first term is straightforward,
\begin{align}
\hspace{-24pt} \frac{\partial f({\tilde{\bf x}},{\tilde{\bf p}}, t)}{\partial t}
= \sum_{nlmNLM} \frac{\partial
a_{nlm}^{NLM}(t)}{\partial t} R_{nl}(\alpha_{nl}, \tilde{r}) Y_{lm}(\hat{r})
&\nonumber\\ \exp(-\tilde{p})
P_N(\tilde{p}) Y_{LM}(\hat{p}) &
\end{align}
Note $\tilde{r}=r/r_{\rm max}$ and $\tilde{p}=p/T_p$. If $r_{\rm max}$ and/or
$T_p$ vary with time, then one needs to
include additional terms which we do not elaborate in details. To obtain the
evolution equation for the expansion
coefficients $a_{nlm}^{NLM}$, we define the following shorthand to project out
the expansion coefficients from any
function $F$,
\begin{eqnarray}
\hspace{-24pt} \langle n'l'm'N'L'M', F({\bf x}, {\bf p}) \rangle = \int_0^{1}
\tilde{r}^2 d\tilde{r} \int d\Omega
\int_0^\infty \tilde{p}^2 d\tilde{p}
 & \ \ \ \nonumber\\
 \int d\Omega_p R_{n'l'}(\alpha_{n'l'},\tilde{r})
 Y^*_{l'm'}(\hat{r}) P_{N'}(\tilde{p}) Y^*_{L'M'}(\hat{p})  F({\bf x}, {\bf p})&
\end{eqnarray}
Performing such projection for the first term, we obtain
\begin{eqnarray}
 \langle n'l'm'N'L'M', \frac{\partial f({\tilde{\bf x}},{\tilde{\bf p}},
t)}{\partial t} \rangle = \frac{\partial a_{n'l'm'}^{N'L'M'}(t)}{\partial t}
\end{eqnarray}
The second term involves the gradient of a function of $r$ times a spherical
harmonics. We note the following gradient
formula,
\begin{eqnarray} \label{gradient_formula}
 \hspace{-24pt} \nabla F(r) Y_{lm}(\hat{r}) = \sum_{i} \hat{\xi}_i
\left[\sqrt{\frac{l}{2l+1}} \left(\frac{dF}{dr} + \frac{l+1}{r} F\right)
 \right. & \ \ \ \nonumber\\ \left. Y_{l-1,m-i}(\hat{r}) (l-1,m-i,1,i|l,m)
  -  \sqrt{\frac{l+1}{2l+1}} \right. &\nonumber\\ \left.   \left(\frac{dF}{dr} -
\frac{l}{r} F\right) Y_{l+1,m-i}(\hat{r}) (l+1,m-i,1,i|l,m)
 \right] &
\end{eqnarray}
Here $(l_1,m_1,l_2,m_2|j,m)$ represent Clebsch-Gordan coefficients  for adding
two angular
momenta ${\bf j} = {\bf l}_1 + {\bf l}_2$. The spherical basis vectors $\xi_i$
are defined as
\begin{align}
 \hat{\xi}_{\pm 1} = \mp \frac{1}{\sqrt{2}} (\hat{e}_x \pm i \hat{e}_y), \, \,
\, \hat{\xi}_0 = \hat{e}_z
\end{align}
The use of spherical basis vectors is convenient as three components of a vector
${\bf V}$ are directly related to spherical harmonics $Y_{1i}$,
\begin{align}
 V_i = |{\rm V}| \sqrt{\frac{4\pi}{3}} Y_{1,i}(\hat{V})
\end{align}
The gradient formula Eq.~(\ref{gradient_formula}) can be further simplified if
one has spherical Bessel function, $F(r) = j_l(k r)$,
with the help of the following recurrence relations
\begin{align}
 \frac{d}{dr} j_l(k r) &= k j_{l-1}(k r) - \frac{l+1}{r} j_l(kr) \nonumber \\
 \frac{d}{dr} j_l(k r) &= -k j_{l+1}(k r) + \frac{l}{r} j_l(kr)
\end{align}
Applying to the first and second terms in Eq.~(\ref{gradient_formula}), one has
for our case
\begin{eqnarray} \label{gradient_formula3}
\hspace{-24pt} \nabla j_{l} (\alpha_{nl} \tilde{r}) Y_{lm}(\hat{r}) =
\frac{\alpha_{nl}}{r_{\rm max}}
 \sum_{i} \hat{\xi}_i  \sum_{\bar{l}} (\delta_{\bar{l},l+1} +
\delta_{\bar{l},l-1})
  & \ \ \ \nonumber\\  \sqrt{\frac{l+\bar{l}+1}{2(2l+1)}} j_{\bar{l}}
(\alpha_{nl} \tilde{r}) Y_{\bar{l},m-i}(\hat{r})
 (\bar{l},m-i,1,i|l,m)  &
\end{eqnarray}
where we have combined two terms together into a compact form. The second term
becomes
\begin{eqnarray}
\hspace{-24pt} {\bf v} \cdot \nabla f({\tilde{\bf x}},{\tilde{\bf p}}, t) = \!\!
\!\! \!\sum_{nlmNLM} a_{nlm}^{NLM}(t)
\frac{\alpha_{nl}}{r_{\rm max}} \frac{\sqrt{2}}{j_{l+1}(\alpha_{nl})} & \ \ \
\nonumber\\ \!\sum_i  \sum_{\bar{l}}
(\delta_{\bar{l},l+1} + \delta_{\bar{l},l-1})
 \sqrt{\frac{l+\bar{l}+1}{2(2l+1)}} j_{\bar{l}} (\alpha_{nl}\tilde{r})
Y_{\bar{l},m-i}(\hat{r})
 & \nonumber\\  (\bar{l},m-i,1,i|l,m)
 \exp({-\tilde{p}}) P_N(\tilde{p}) Y_{LM}(\hat{p}) \sqrt{\frac{4\pi}{3}}
Y_{1i}(\hat{p}) &
\end{eqnarray}
Following the same procedure as done for the first term, we project out the
expansion coefficients $a_{nlm}^{NLM}$ for the second
term,
\begin{eqnarray}
\hspace{-24pt} \langle n'l'm'N'L'M', {\bf v}  \!\cdot\! \nabla f({\tilde{\bf
x}},{\tilde{\bf p}}, t) \rangle = \!\!
\!\! \!\sum_{nlmNLM} \!\sum_i a_{nlm}^{NLM}(t)
& \ \ \ \nonumber\\
\frac{\alpha_{nl}}{r_{\rm max}}
 \!\! \sum_{\bar{l}} (\delta_{\bar{l},l+1} + \delta_{\bar{l},l-1})
 \sqrt{\frac{l+\bar{l}+1}{2(2l+1)}} (\bar{l},m-i,1,i|l,m) & \nonumber\\
  \frac{\sqrt{2}}{j_{l'+1}(\alpha_{n'l'})} \frac{\sqrt{2}}{j_{l+1}(\alpha_{nl})}
   \int_0^{1} \!\! \tilde{r}^2 d\tilde{r} j_{l'}(\alpha_{n'l'}\tilde{r})
j_{\bar{l}} (\alpha_{nl} \tilde{r})
 & \nonumber \\  \int \!\!d\Omega  Y^*_{l'm'}(\hat{r}) Y_{\bar{l},m-i}(\hat{r})
\! \int_0^\infty \!\! \!\tilde{p}^2 d\tilde{p}   P_{N'}(\tilde{p}) 
\exp({-\tilde{p}}) P_N(\tilde{p}) & \nonumber\\ \int
\!\!d\Omega_p Y^*_{L'M'}(\hat{p})Y_{LM}(\hat{p}) \! \sqrt{\frac{4\pi}{3}}
Y_{1i}(\hat{p})  \ \ &
\end{eqnarray}
The integrals $\int d\Omega$ and $\int d\tilde{p}$  can be done using orthogonal
relations for spherical harmonics and
Laguerre function. The integral $\int d\Omega_p$ involves the product of three
spherical harmonics, which can
performed with the help of the following relation,
\begin{eqnarray}
 \hspace{-24pt} \int d\Omega_p  Y^*_{L'M'} (\hat{p})Y_{LM}(\hat{p})
Y_{1i}(\hat{p})
 = \sqrt{\frac{3(2L+1)}{4\pi(2L'+1)}} & \nonumber \\ (L,0,1,0|L',0)
 (L,M,1,i|L',M') &
\end{eqnarray}
Note that the above Clebsch-Gordan coefficients are non-zero only when $|L-L'|
\le 1$ and $L - L' - 1$ are
even numbers. This implies that $L = L'\pm 1$ and $M = M'-i$. The final result
for the second term is,
\begin{eqnarray}
\hspace{-24pt} \langle n'l'm'N'L'M', {\bf v}  \!\cdot\! \nabla f({\tilde{\bf
x}},{\tilde{\bf p}}, t) \rangle = \!\!
\!\!\! \sum_{nlmNLM} \! \sum_i a_{nlm}^{NLM}(t)
& \ \ \ \nonumber\\
 \frac{\alpha_{nl}}{r_{\rm max}} I_r[n',l',n,l] \delta_{NN'}
(\delta_{l',l+1} + \delta_{l',l-1}) \sqrt{\frac{l+l'+1}{2(2l+1)}} &\nonumber\\
(\delta_{L',L+1} + \delta_{L',L-1}) \sqrt{\frac{2L+1}{2L'+1}} (L,0,1,0|L',0)
&\nonumber\\
 \delta_{m',m-i}  (l',m-i,1,i|l,m) \delta_{M',M+i}  (L,M,1,i|L',M')
&\nonumber\\ 
\end{eqnarray}
where $I_r[n',l',n,l]$ has been defined in Eq.~(\ref{Ir_nlnl}). Combining with
the first term, we finish the multipole
expansion of the left hand side in Boltzmann equation.


\bibliographystyle{h-physrev5.bst}
\bibliography{GYQ_refs.bib}
\end{document}